# Design space for low sensitivity to size variations in [110] PMOS nanowire devices: The implications of anisotropy in the quantization mass

Neophytos Neophytou and Gerhard Klimeck*

Network for Computational Nanotechnology, Birk Nanotechnology Center, Purdue University, West Lafayette, Indiana 47907-1285

*Corresponding author's email: gekco@purdue.edu

## ABSTRACT

A 20-band $sp^3d^5s^*$ spin-orbit-coupled, semi-empirical, atomistic tight-binding model is used with a semi-classical, ballistic, field-effect-transistor (FET) model, to examine the ON-current variations to size variations of [110] oriented PMOS nanowire devices. Infinitely long, uniform, rectangular nanowires of side dimensions from 3nm to 12nm are examined and significantly different behavior in width vs. height variations are identified and explained. Design regions are identified, which show minor ON-current variations to significant width variations that might occur due to lack of line width control. Regions which show large ON-current variations to small height variations are also identified. The considerations of the full band model here show that ON-current doubling can be observed in the ON-state at the onset of volume inversion to surface inversion transport caused by structural side size variations. Strain engineering can smooth out or tune such sensitivities to size variations. The cause of variations described is the structural quantization behavior of the nanowires, which provide an additional variation mechanism to any other ON-current variations such as surface roughness, phonon scattering etc.

**Index terms –** **nanowire, bandstructure, tight binding, transistors, MOSFETs, variations, effective mass, injection velocity, quantum capacitance, anisotropy.**

*Motivation:* As transistor sizes shrink down to the nanoscale, a possible device approach that has attracted large attention recently because of its possibility of enhanced electrostatic control, is the multi-gated nanowire (NW) transistor [1]. Nanowire transistors of diameters even down to 3nm have already been demonstrated by various experimental groups [2-6]. At such small scales, however, the issue of device sensitivity to parameter fluctuations will be critical. Atomistic variations of the side lengths, surface roughness, line edge roughness, cross section shape variations, defects, surface states will exist in these devices and need to be tolerated (if at all possible). Device orientation as well as the quantization surfaces will also be an important design parameter. In the case of nanowires (and thin body devices), the high symmetry orientations [100], [110] and [111] as shown in Fig. 1, have been extensively studied. Both experiments and simulations have identified that for NMOS nanowires the beneficial transport orientations are [110] and [100] [7-9] to deliver the highest currents. In the case of ballistic PMOS devices, however, simulation has shown that the [100] transport orientation lacks behind the [110] orientation [10, 11] in its current currying capabilities. For this reason, and because of the fact that the optimized conventional CMOS architecture orientations are (001)/[110], it would be beneficial to build devices on the (001) surface and utilize the [110] transport direction. The sensitivity of that geometry (Fig. 1b), to size fluctuations, is the subject to detailed discussions. Although several reasons might add to the device performance variations such as interface roughness and phonon scattering, this work considers infinitely uniform rectangular nanowires, and investigates the sensitivity of ON-current to side size variations, resulting alone from the internal structural and electrostatic quantization behavior of the nanowire. These fluctuations are attributed to the anisotropic hole effective mass, and cannot be captured appropriately in effective mass models. This additional mechanism can result in current fluctuations of 100% while typical treatments of surface roughness scattering at ON-state have resulted in rather modest current variations of 10-20% [12, 13, 14]. The electronic structure effect discussed and explained here is an additional variation mechanism on top of the already existing mechanisms, and we believe it is significantly stronger than the perturbative effects typically considered in surface roughness models.

*Summary of the paper:* In this work, an atomistic nearest-neighbor tight binding (TB) model ($sp^3d^5s^*$-SO) [15-18] is used for the nanowires' electronic structure calculation, coupled to a 2D Poisson solver for electrostatics. To evaluate transport characteristics, a simple semi-classical ballistic model [19, 20] is used. The variations in the ON-current with size variation of [110] PMOS rectangular nanowire devices with (1-10) and (001) quantization surfaces, as shown in Fig. 1b, are investigated. Cross sectional widths and heights with lengths from 3nm to 12nm (all combinations of side lengths) are considered. The nanowires are considered infinitely long in the transport orientation, with uniform surfaces. Design regions in which the current is at large extent tolerant to the nanowires' side variations are identified. Fluctuations in the [1-10] direction (width size, or equivalently fluctuations in the (001) surface area) have a small and almost linearly varying impact on ON-current variations. [001] (height) length variation (equivalently variations in the (1-10) surface area) has a more complicated impact on the trend of the ON-current variation, with large ON-current sensitivity for nanowires with [001] heights of lengths 6nm-8nm. This behavior appears at the onset of volume inversion to two surface inversion channels. Its specific side length appearance (6nm-8nm) originates in the internal structural quantization and electrostatic confinement behavior of the nanowires. The reason it is only observed in the [001] direction is a result of the anisotropy of the Si heavy-hole (HH) valence band, which strongly affects the preference of charge placement in the wires' cross section and along its quantization surfaces. It is also observed for different gate biases at very similar side lengths. It is shown that strain engineering can change the anisotropy of the heavy-hole subband and make the sensitivity of the ON-current to side variations more uniform, or tune the sensitivity to different design regions.

*Necessity of atomistic modeling:* The problem of identifying the correct bandstructure for the valence band of Si in the inversion layers is complicated (especially for nanowires), because of the strong non-parabolicity and anisotropy of the heavy-hole and its coupling to the light-hole (LH). Several authors have investigated various techniques for description of the valence band [21-23]**,** for both unstrained and strained

MOSFET channels. In addition, under extreme scaling of device dimensions, the atoms in the cross section will be countable, and crystal symmetry, bond orientation, distortions, surface truncation, and quantum mechanical confinement will dominate transport characteristics [7, 10]. The nearest neighbor TB $sp^3d^5s^*$-SO model used in this work, with a basis set composed of localized orthogonal orbitals, is most appropriate for this purpose since it inherently includes all of the above features. The model itself and the parameterization presented in [15] have been extensively calibrated to various experimental data of various nature with excellent agreement (details and references in [7]).

*The simulation approach:* The devices simulated are rectangular nanowires in the [110] transport orientations with 1.1nm $SiO_2$ oxide thickness. [001] and [1-10] are the two equivalent quantization directions (Fig. 1b). The simulation procedure consists of three steps as described in detail in [7] and summarized here:
1. The bandstructure of the wire is calculated using the $sp^3d^5s^*$-SO model. The atoms that reside on the surface of the nanowire are passivated in the $sp^3$ hybridization scheme [24].
2. A semi-classical top-of-the-barrier ballistic model is used to fill the dispersion states and compute the transport characteristics [19, 20].
3. A 2D Poisson equation is solved in the cross section of the wire to obtain the electrostatic potential. The electrostatic potential is added to the diagonal on-site elements of the atomistic Hamiltonian as an effective potential for recalculating the bandstructure until self consistency is achieved.

Although the transport model used is a simple ballistic model, it allows for examining how the bandstructure of the nanowire alone will affect its ballistic transport characteristics. The same conclusion to parts of this work can be obtained from full 3D quantum (NEGF) simulations [11, 25, 26, 27], and still simulations might be restricted to smaller nanowire cross sections (rather than up to the 12nm x 12nm cross sections we are considering). The simple model used here, however, provides critical physical insight. It is the simplicity of the transport model, which allows to shed light on the importance of the dispersion details and the charge distributions, which might get lost in a full-fledged quantum transport simulation. The results presented in this work, focus on the ON-current variation behavior of PMOS [110] nanowires. More detailed transport properties of PMOS nanowires, also in different transport orientations, are presented in [10].

*Valence band anisotropy affects quantization:* It is well known that the valence bands of the standard semiconductors are very anisotropic with a general rule of thumb of $m_{[100]} < m_{[110]} < m_{[111]}$ for the heavy-hole. For Si we find $m_{[110]}$ =-0.579 and $m_{[100]}$ = -0.275, therefore $m_{[110]} / m_{[100]} \sim 2.1$, which is a very significant distortion. The light-hole bands show typically significant less anisotropy with $m_{[110]}$ = -0.147 and $m_{[100]}$ = -0.204, therefore $m_{[110]} / m_{[100]} \sim 0.72$. With these mass values a simple particle in a box model predicts an energy separation of *dE=0.26eV* for the heavy-hole and light-hole ground states in a 3nm 2D-box. The ground state, therefore, in a PMOS nanowire is dominated by the strongly anisotropic heavy-hole states. This argument will be used in the semi-analytical explanation of the dispersion and quantization behavior. However, the nanowire dispersions we compute include all bands including heavy-, light-hole and split-off.

Figure 2a shows the (001) surface energy contour of the bulk heavy-hole Si valence band. The anisotropy is clearly evident in the bandstructure between the [100] and [110] directions. Fig. 2b shows the (110) surface contour (the plane perpendicular at -45° in Fig. 1a). The [10-1] and [100] directions indicated in this figure are the relevant quantization directions of the [110] oriented structure examined in this work. The elongation along the [1-10] direction (Fig. 2b, d) indicates a heavier quantization mass than in the [001] direction, which makes the valence band edge more sensitive to variations in the [001] side (smaller mass) than in the [1-01] side (larger mass). Figure 2c shows the band edge of [110] nanowire devices (as in Fig. 1b) starting from a 3nm x 3nm wire, and increasing the width, [1-10], or the height, [001], directions to 9nm. Changes in the width, [1-10] (black), cause smaller variations to the band edge compared to changes in the height, [001] (blue). Most of the total band edge shift because of 2D quantization is a result of the lighter quantization mass (001) surface, in agreement with the particle in a box quantization picture.

*Anisotropy implications on device performance:* In that scope, Kobayashi *et al.* in [5], showed experimentally that different quantizations impact the performance of nanowire devices through $V_T$ fluctuations and ON-current variations directly originating from the anisotropic band variation. In that work, it was shown that $V_T$ and $I_{ON}$ of PMOS nanowire devices are very sensitive to [100] side variations, but much less sensitive to [110] side variations. This is evidence of the heavier [110] mass quantization that does not allow large subband variations with size fluctuations.

*Different charge distribution in different orientations:* The anisotropy in the bandstructure also affects the charge distribution in the cross section in the wire. As we have shown earlier [10], in the case of the [110] wires quantized in the [001] and [1-10] directions, the charge tends to accumulate closer to the heavy quantization mass (1-10) surfaces rather than the lighter (001) ones. This is also shown in the smaller sub-figures surrounding Fig. 3a, that show device cross sections and the charge distribution under high bias conditions. The top/bottom surfaces in these figures are (001), whereas the left/right ones are (1-10) surfaces. Figure 3a (central) shows the ON-current of the nanowires as a function of their height- (in the [001] direction) and their width- (in the [1-10] direction) as they change from 3nm to 12nm. More on the details of the centered figure will be discussed further on.

The charge distribution in sub-Fig. 3a(iii-v) in the bottom row, for widths [1-10] 3nm, 6nm and 12nm respectively, shows that the charge is preferably accumulated on the (1-10) left/right surfaces in agreement with ref [28] and splits into two lobes as the width increases. In the body of the wire, as well as along the (001) top/bottom surfaces, smaller charge accumulation is observed. The situation is different in the case were the height-[001] increases from 3nm to 12nm in sub-Fig. 3a(iii, ii, i) shown in the left column. The charge is accumulated closer to the left/right (1-10) surfaces, and finally two parallel "3nm x 3nm wire" like channels are formed at the top/bottom regions of the nanowire. As the dimensions of the device increase to 6nm x 12nm and 12nm x 12nm (sub-Fig. 3a(vi, vii) respectively, in the right column), clearly, a stronger inversion layer is formed along the (1-10) surfaces (right/left), rather than the (001) surfaces (top/bottom). The inversion layer on the (1-10) surface extents ~1.5nm. This distance almost doubles in the case of the (001) surface. We would like to mention here that the charge placement in the devices' corners is a pure electrostatic effect coming from the stronger inversion near the corners of the device due to stronger electric fields. The electrostatic potential in the width and height directions, however, is virtually identical as shown in ref [10]. The charge along the surface is the quantity that depends on the quantization mass and the detailed crystal symmetry.

*Implications on the ON-current variations with size variations:* The centered plot of Fig. 3a shows the ballistic ON-current of [110] oriented nanowire devices as a function of the *x*-axis-width-[1-10] and *y*-axis-height-[001] dimensions of the device. All width/height combinations from 3nm x 3nm of nanowires (left/bottom corner), all the way to 12nm x 12nm wires (right/upper corner) are presented in steps of 1nm. (Around the 6nm-8nm height size, the steps used are 0.5nm). All parameters in the simulation are fixed in all cases, with only the dimensions changing. The gate bias is set to $V_G$=1V, drain bias $V_D$=0.5V in all cases, and the insulator thickness $t_{ins}$=1.1nm. The current plotted is in μA, while contour lines are drawn every 5μA.

*Clear boundary identified between two insensitive regions:* Starting from the 3nm x 3nm wire (left/bottom corner) where the current is the lowest, the current levels increase as the dimensions of the device increase to 12nm x 12nm (right/upper corner). Regions where the ON-current does not significantly vary with size variations and others that suffer from enhanced variations with size variations can be identified. A region very lightly affected by size variations is the one between 3nm – 6nm of height, and for any width (region A). Within this region, the current does not vary significantly with changing width. Increasing the width from 3nm to 12nm (300%), and equivalently the perimeter by 150%, only increases the ON-current by 50%. The region from 8nm-12nm of height and any width (region B) can also be considered to be relatively tolerant to height variations, although somewhat higher ON-current variation is observed at widths of 10nm-12nm. This variation behavior is almost linear with size variations. In contrast, the region between 6nm-8nm of height and for any width shows very sharp ON-current variations with relatively small height variations, and needs to be avoided for a design to be tolerant to variations. A clear boundary between two regions that can be considered relatively insensitive to variations is therefore identified. This current variation originates from the structural and electrostatic quantization behavior of nanowires' bandstructure in the transverse direction. It is also observed in lower gate biases ($V_G$=0.5V), although somewhat smoothed out, i.e. the fast current varying region is expanded to heights-[100] of 6nm-9nm. Same qualitative results above were also obtained by using a different set of TB parameters obtained from [29], with the same fast varying current region between 6-8nm.

*The ON-current variation with width-[1-10] variation*: The shape of the charge distribution in the cross section of the device as the width increases sheds light on the reason that large variations in the width [1-10] direction do not result in large variations in the ON-current. As shown in sub-Fig. 3a(iii, iv, v), at the bottom of Fig. 3a, the charge has formed two channels, on the left/right of the channel. Increasing the width (at a constant 3nm height), is equivalent to increasing the upper-lower (001) surface areas. Hence, the (001) surface current, as well as the current in the middle of the wire, both increase. This causes a controllable and almost linear change in the ON-current as the width changes (at a constant height). In region labeled "B", for devices with heights from 8nm-12nm, as the width changes, the changes in the ON-current come from changes in the upper//lower surface areas at the top/ bottom of the nanowire (sub-Fig. 3a(vii)). Similar ON-current variations are therefore observed.

*The ON-current variation with height, [001], side variation*: Variations in the [001] equivalent quantization direction also do not cause significant variations in the ON-current performance, except in the region between 6nm-8nm of width, in which the variations are very large (the ON-current almost doubles with only 2nm increase in the width). The explanation is also understood from the charge distribution sub-figures along the left/right of Fig. 3a. As the [001] quantization height increases, at some point around 6nm, the charge distribution splits into two "3nm-like" wires on the top/bottom (001) equivalent surfaces. Two channels are formed now. The ON-current undergoes a sharp increase during this formation. (The reason this sudden change is not observed in the case of width increase is that the two channels have already been formed at the 3nm width due to the heavier [1-10] direction quantization mass). Further increase in the height up to 12nm (at any constant width-[1-10]), increases the length of the inversion layer charge along the left/right (1-10) equivalent surfaces. The ON-current however does not follow a smooth and linear increase, but it is rather not-sensitive to variations with small oscillations observed. This has to do with the interplay between the carrier velocity and charge as the height-[001] increases as it will be explained further down.

*Charge variations:* Figure 3b shows the charge variations corresponding to the ON-currents presented in Fig. 3a. The charge variation is very symmetric, with respect to the width and height of the device. (A purely symmetric case would be a mirror image about the -45° line across the figure). It seems that the charge of devices with same perimeter length is very similar, independently if the largest surface is (1-10) or (001). This is an observation also noted in [7] and [30], when comparing channels of materials with different quantum capacitance ($C_Q$). In these works, it was shown that in channels with bias dependent quantum wells, differences in $C_Q$ are smeared out and create much less differences in the total gate capacitance and the inversion layer charge of the device.

*Velocity variations:* In the ballistic limit, the ON-current can be calculated by the product of charge times the average carrier velocity. Since the charge is very similar for devices with the same perimeter, the significant differences in the ON-current must originate from the velocity differences. In Fig. 4a the average velocity of the carriers is plotted. (This is defined as the total current divided by the total charge in each of the nanowires). The velocity contour plot in Fig. 4a has indeed a very different pattern than the charge pattern, and it is the major reason behind the ON-current variation pattern. Noticeable in this figure is the three almost similar velocity regions that can be identified for heights-[001] from 3-6nm, ~6-9nm and ~9-12nm and for any width-[1-10]. In Fig. 4(b-f) the dispersions of the nanowires for the devices labeled (b-f) in Fig. 4a are shown. There are two main features that can be identified in the dispersions. The first one is the light mass subbands at higher energies, and the second is the heavier subbands at lower energies, near or below the Fermi level. (*Efs* denotes the position of the Fermi level at the particular bias point). We would like to stress out here that these light/heavy subbands should *not* be identified as been heavy or light-hole like subbands. They are rather mixture of the two. As explained in ref [10], the light subbands originate from the quantization of the heavy-hole in the [1-10] quantization direction, from states that are physically or electrostatically quantized (and not from the light-hole). The heavy subbands originate mostly from the heavy hole states that are lower in energy (outside the quantized potential well) and do not feel large quantization.

*Light subbands along (1-10) surfaces - increase in the height-[001]:* For a constant width-[1-10] and varying height-[001] as labeled (b), (c) and (d) in Fig. 4a, the number of occupied *light* subbands increases from 4 to 8 and then to 12. The charge distribution from these light subbands (with energies near the ground state) accumulates in the potential wells formed along the inverted left/right (1-10) surfaces. Increasing the area of these surfaces, increases the number of lighter, near ground state subbands. An increase in the number of the lighter subbands indicates an increase in the average carrier velocities. Indeed, the average carrier velocity increases as the height-[001] increases from 3nm to 9nm. As the height of the nanowire increases more, however, the number of the heavier subbands (with energies farther from the ground state and wavefunctions more spread in the body of the wire) also increases, and the average velocity reduces (Fig. 4a, region (d)). An interplay between the light and heavy subbands, is what determines the carrier velocity.

*Heavier subbands along (001) surfaces - increase in the width-[1-10]:* In the horizontal direction, on the other hand, as the width-[1-10] increases, the (001) surface increases (labels (e), (c) (f) in Fig. 4a). The inversion charge on the (001) surfaces resides further in the nanowires' body than in the (1-10) surface charge, and is primarily composed of the heavier transport mass subbands. The number of light subbands in the dispersions in Fig. 4e, c, f, remains constant at 8 subbands in all cases, but the number of heavy subbands increases as the width increases, and the average velocity drops as the wire widths change from region (e) to (c) and finally to (f). The interplay between the charge and velocity surfaces results in the current surface shown in Fig. 3a.

*On the additional role of surface roughness scattering (SRS) on device variations:* Having examined this internal to the nanowire properties quantization behavior and its implication on ON-current variations, we would like to stress that the mechanism described in Fig. 3, at which the current undergoes a large increase as the internal charge placement undergoes a transition from two to four surface channels is a fundamental one, solely determined by the crystal direction and the anisotropic masses. Surface roughness will surely modulate the internal mode spectrum and cause performance reduction due to mode-to-mode scattering. However (SRS) will not eliminate the formation of 2 to 4 mode transition which is rather evident here. The effective doubling of the channels that is sensitive to the height variation, but not the width variation roughly causes a doubling of the current (~100% increase), which is much larger than any simulated SRS variation effects in similar cross section size nanowires (~10%-20% at high inversion conditions) [12, 13, 14].

For small side length wires (<4nm), SR causes performance variations through deforming the subbands by creating local barriers and wells [14, 31]. A theoretical quantum transport (NEGF) study in [14], showed that SR for wires of side lengths of 3 nm suffer from mobility degradation and large variations at low gate biases, however, the performance is partly recovered and variations in the performance are reduced at high inversion conditions. The reduced effect of SRS in narrow nanowires at high gate biases was also concluded from a semiclassical study in [13]. In another quantum transport study, Wang *et* al. [12], showed that the ballisticity of a 3nm x 3nm rough nanowire was close to 85%, which does not leave much space for significant variations due to SRS. The reason for the reduced effect of SR variations at high bias, is that as the gate bias increases and more carriers are in the channel, further propagating states appear, localization effects become less important, and the potential wells/barriers are smoothen/widen out.

In the case of wires with larger sides (>5nm), where more modes are now occupied, SRS affects the device mostly through mode coupling/mixing as shown in full 3D transport in an effective mass model [14]. Poli *et* al. [14], showed that for 20nm long nanowires of 5nm x 5nm, and 7nm x 7nm cross sections, SRS itself, can only degrade the mobility of the nanowire by ~10% for both low and high gate biases. In that work, statistics on 20 different roughened nanowire samples, showed that the variation in the characteristics of the samples was less than ~10%. (Larger devices feel the effect of averaging more, and variations are reduced, and since the mobility of the roughened devices is very close to the mobility of the ideal devices (~90%), there is no room for large variations). We plan to examine these conclusions with our full 3D atomistic transport model in more detail, especially to examine the effects of atomistic disorder. We believe, however, that the fundamental conclusion of channel formation and different height and width sensitivities govern the ON-current transport. Fluctuations along the channel will no doubt affect the performance and introduce some variations that will modulate the current further, but the different sensitivity in height and width will remain.

*Strain engineering to tune the sharp current variation regions:* In the case of PMOS devices, uniaxial compressive strain engineering has been utilized to enhance performance [32]. Here, the effect of two different strain tensor cases on the current variations is examined: One that reduces, and one that enhances the heavy-hole anisotropy.

*Reduced anisotropy:* Introducing 3% compressive strain in the transport orientation, 3% tensile in the [1-10] quantization orientation and moderate compressive strain (0.05%) in the [001] quantization orientation, can make the quantization surface to look almost isotropic at least for higher energies (Fig 5a), while still having a light transport mass. In this case, as shown in Fig 5b, at the expense of losing some of the tolerance to variations in the previously more insensitive regions, the fast varying boundary can be almost completely removed for widths-[1-10] below 8nm. It still appears slightly in the 8nm-12nm height-[001] region. The price of this however, is the loss of the large variation tolerance in the rest of the design space.

*Enhanced anisotropy case:* On the other hand, introducing a different strain tensor (in this case 3% compressive strain in the two quantization directions and 1% compressive strain in the transport direction), enhances the anisotropy of the heavy-hole as shown in Fig. 5c. This causes the sharp varying region to shift to larger [001] heights, around 9nm as shown in Fig. 5d. A reduction in the [001] quantization mass, allows a larger spread for the wavefunction in the [001] direction of the wire's cross section, which shifts the transition between a single to double channel at larger [001]-heights. (We mention here that this strain combination increases the transport effective mass, so the current levels are lower, however, it is just a demonstration on how the insensitivity to side size variations can be tuned with strain engineering).

In summary, the effect of side length sensitivity in the ballistic transport properties of infinitely long and uniform PMOS [110] oriented nanowires with width-[1-10] / height-[001] dimensions from 3nm up to 12nm was examined. The [110] wires examined, with (1-10)/(001) quantization surfaces, have asymmetric charge distribution in their cross section, with preferable accumulation along the (1-10) surface which has a higher quantization mass. Variations in the [1-10] wire width cause only small and linear variations in the ON-current. Variation in the [001] wire height appears to have large impact in on the ON-current variation around the 6nm-8nm length region, where the transport shifts from volume inversion to two surface inversion layer transport on the two (001) surfaces (equivalently, from two to four lobes, one in each corner). This effect will appear in any situation at which the device shifts from bulk/volume-like transport to two surface-like transport channels. The placement of the boundary in that respect will depend on the quantization masses. Strain engineering can smoothen out the large variation of the ON-current, or can tune the sensitivity to different design regions. These observations can give guidance towards the design of multi-surface devices such as nanowires and FinFETs.

The authors would like to mention that the simulator used in this study will be released as an enhanced version of the Bandstructure Lab on nanoHUB.org [33]. This simulation engine will allow any user to duplicate the simulation results presented here. Over 1,800 users have run over 12,000 simulations in the existing Bandstructure Lab, which has not yet included the charge self-consistent transport model we demonstrate here. This new charge self-consistent capability has been added very recently (August 2008).

# Acknowledgements

This work was funded by the Semiconductor Research Corporation (SRC). The computational resources for this work were provided through nanoHUB.org by the Network for Computational Nanotechnology (NCN). The authors would like to acknowledge Prof. Timothy Boykin of University of Alabama at Huntsville for tight-binding discussions.

# References

[1]  ITRS Public Home Page. http://www.itrs.net/reports.html

[2] N. Singh, F. Y. Lim, W. W. Fang, S. C. Rustagi, L. K. Bera, A. Agarwal, C. H. Tung, K. M. Hoe, S. R. Omampuliyur, D. Tripathi, A. O. Adeyeye, G. Q. Lo, N. Balasubramanian, D. L. Kwong, "Ultra-narrow silicon nanowire gate-all-around CMOS devices: Impact of diameter, channel-orientation and low temperature on device performance," *Int. Elec. Dev. Meeting*, 2006.

[3] K. H.Cho, Y. C. Jung, B. H. Hong, S. W. Hwang, J. H. Oh, D. Ahn, S. D. Suk, K. H. Yeo, D.-W. Kim, D. Park, W.-S. Lee, "Observation of single electron tunneling and ballistic transport in twin silicon nanowire MOSFETs (TSNWFETs) fabricated by top-down CMOS process," *Int. Elec. Dev. Meeting*, 2006.

[4] K. H. Cho, K.H. Yeo, Y.Y Yeoh, S. D. Suk, M. Li, J. M. Lee, M.-S. Kim, D.-W. Kim, D. Park, B. H. Hong, Y. C. Jung, and S. W. Hwang, "Experimental evidence of ballistic transport in cylindrical gate-all-around twin silicon nanowire metal-oxide-semiconductor field-effect transistors," *Appl. Phys. Lett.*, 92, 052102, 2008.

[5] M. Kobayashi and T. Hiramoto, "Experimental study on quantum confinement effects in silicon nanowire metal-oxide-semiconductor field-effect transistors and single-electron transistors," *J. Appl. Phys.*, 103, 053709, 2008.

[6] J. Xiang, W. Lu, Y. Hu, Y. Wu, H. Yan, and Charles M. Lieber, "Ge/Si nanowire heterostructures as high-performance field-effect transistors," *Nature*, vol. 441, no. 25, 2006.

[7] N. Neophytou, A. Paul, M. Lundstrom and G. Klimeck, "Bandstructure effects in silicon nanowire electron transport," IEEE Trans. Electron Devices, vol. 55, no. 6, pp. 1286-1297, 2008.

[8] M. Luisier, A. Schenk, W. Fichtner, "Full-Band Atomistic Study of Source-To-Drain Tunneling in Si Nanowire Transistors," Proc. of SISPAD, 978-3-211-72860-4, pp. 221-224, 2007.

[9] M. Yang *et* al., "Hybrid-orientation technology (HOT): opportunities and challenges," *IEEE Trans. Electron Devices*, vol. 53, no. 5, pp. 965-978, 2006.

[10] N. Neophytou, A. Paul, and G. Klimeck, "Bandstructure effects in silicon nanowire hole transport," *IEEE Trans. Nanotechnol.*, vol. 7, issue. 6, pp. 710-719, 2008.

[11] M. Luisier and A. Schenk, "Atomistic simulation of nanowire transistors", *Journal of Computational and Theoretical Nanoscience,* vol. 5, pp. 1-15, June 2008.

[12] J. Wang, E. Polizzi, A. Ghosh, S. Datta, and M. Lundstrom, "Theoretical investigation of surface roughness scattering in silicon nanowire transistors," *Appl. Phys. Lett.*, 87, 043101, 2005.

[13] M. Lenzi, A. Gnudi, S. Reggiani, E. Gnani, M. Rudan, and G. Baccarani, "Semiclassical transport in silicon nanowire FETs including surface roughness," *J. Comput. Electron.* 7: 355-358, 2008.

[14] S. Poli, M. G. Pala, T. Poiroux, S. Deleonibus, and G. Baccarani, "Size dependence of surface-roughness-limited mobility in silicon-nanowire FETs," *IEEE Trans. Electr. Dev.*, vol 55, no. 11, 2008.

[15] T. B. Boykin, G. Klimeck, and F. Oyafuso, "Valence band effective-mass expressions in the $sp^3d^5s^*$ empirical tight-binding model applied to a Si and Ge parametrization," *Phys. Rev. B*, vol. 69, pp. 115201-115210, 2004.

[16] G. Klimeck, F. Oyafuso, T. B. Boykin, R. C. Bowen, and P. von Allmen, *Computer Modeling in Engineering and Science (CMES)*, vol. 3, no. 5, pp. 601-642, 2002.

[17] G. Klimeck, S. Ahmed, H. Bae, N. Kharche, S. Clark, B. Haley, S. Lee, M. Naumov, H. Ryu, F. Saied, M. Prada, M. Korkusinski, and T. B. Boykin, "Atomistic simulation of realistically sized nanodevices using NEMO 3-D—Part I: Models and benchmarks," *IEEE Trans. Elecron Devices*, vol. 54, no. 9, pp. 2079-2089, 2007.

[18] J. C. Slater and G. F. Koster, "Simplified LCAO method for the periodic potential problem," Phys. Rev., vol 94, no. 6, 1954.

[19] M.S. Lundstrom, and J. Guo, "Nanoscale transistors: Device physics, modeling and simulation," *Springer*, 2006.

[20] A. Rahman, J. Guo, S. Datta, and M. Lundstrom, "Theory of ballistic nanotransistors," *IEEE Trans. Electron Devices*, vol. 50, no. 9, pp. 1853-1864, 2003.

[21] M. V. Fischetti, Z. Ren, P. M. Solomon, M. Yang, and K. Rim "Six-band k.p calculation of the hole mobility in silicon inversion layers: Dependence on the surface orientation, strain, and silicon thickness," *J. Appl. Phys.,* vol. 94, no. 2, 2003.

[22] E. X. Wang, P. Matagne, L. Shifren, B. Obradovic, R. Kotlyar, S. Cea, M. Stettler, and M. Giles, "Physics of hole transport in strained silicon MOSFET inversion layers," *IEEE Trans. Electron Devices,* vol. 53, no. 8, 2006.

[23] M. De Michelis, D. Esseni, Y. L. Tsang, P. Palestri, L. Selmi, A. G. O'Neil, and S. Chattopadhyay, *IEEE Trans. Electron Devices,* vol. 54, no. 9, 2007.

[24] S. Lee, F. Oyafuso, P. Von, Allmen, and G. Klimeck, "Boundary conditions for the electronic structure of finite-extent embetted semiconductor nanostructures," *Phys. Rev. B*, vol. 69, pp. 045316-045323, 2004.

[25] M. Luisier, and G. Klimeck, "Full-band and atomistic simulation of n- and p-doped double-gate MOSFETs for the 22nm technology node," *Simulation of Semiconductor Processes and Devices,* 2008. SISPAD 2008. International Conference on 9-11 Sept. 2008 Page(s):17 - 20 Digital Object Identifier 10.1109/SISPAD.2008.4648226, IEEE Proceedings

[26] M. Luisier, and G. Klimeck, "OMEN an Atomistic and Full-Band Quantum Transport Simulator for post-CMOS Nanodevices," *8th IEEE Conference on Nanotechnology*, 2008. NANO '08., 18-21 Aug. 2008 Page(s):354 – 357, Digital Object Identifier 10.1109/NANO.2008.110, IEEE proceedings

[27] M. Luisier, N. Neophytou, N. Kharche, and G. Klimeck, "Full-Band and Atomistic Simulation of Realistic 40 nm InAs HEMT", *in Proc. of IEDM*, San Francisco, USA, Dec. 15-17, 2008.

[28] M. Saitoh, S. Kobayashi, and K. Uchida, "Physical Understanding of Fundamental Properties of Si (110) pMOSFETs Inversion-Layer Capacitance, Mobility Universality, and Uniaxial Stress Effects," *in Proc. of IEDM,* 2007, pp. 711-714.

[29] J. M. Jancu, R. Scholz, F. Beltram, F. Bassani, "Empirical spds* tight-binding calculation for cubic semiconductors: general method and material parameters," *Phys. Rev. B*, vol. 57, no. 11, 15, 1998.

[30] H. S. Pal, K. D. Cantley, S. S. Ahmed, and M. S. Lundstrom, "Influence of bandstructure and channel structure on the inversion layer capacitance of silicon and GaAs MOSFETs," IEEE Trans. Electron Dev., vol. 55, pp. 904-908, 2008.

[31] K. Uchida, and S. Takagi, "Carrier scattering induced by thickness fluctuation of silicon-on-insulator film in ultrathin-body metal-oxide-semiconductor field-effect transistors," *Appl. Phys. Lett.*, vol. 82, no. 17, 2003.

[32] S.E. Thompson et al., "A 90-nm logic technology featuring strained-silicon," *IEEE Trans. Electron Devices,* vol. 51, no. 11, pp. 1790-1797, 2004.

[33] [nanoHub] Bandstructure lab on nanoHUB.org (https://www.nanohub.org/tools/bandstrlab/)

Figure 1: Wires in different orientations

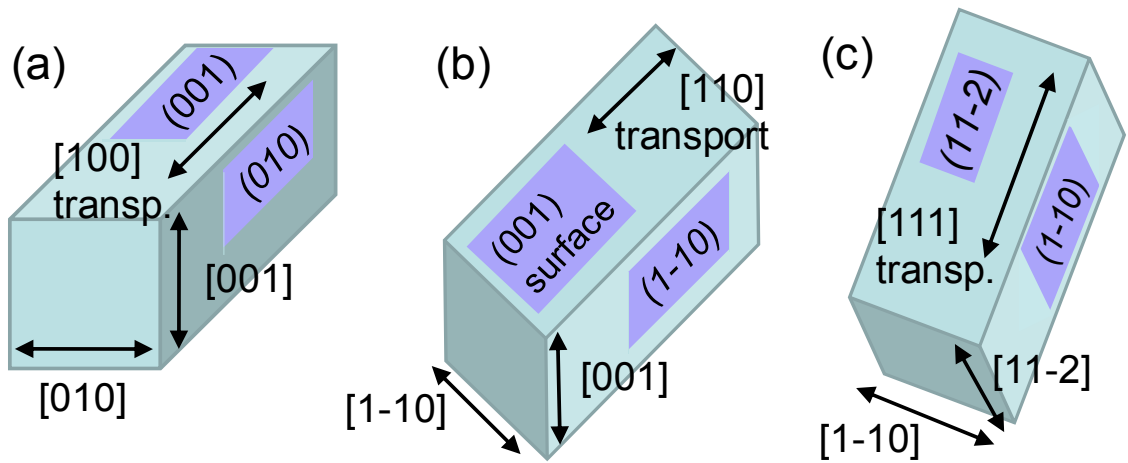

Figure 1 caption:

Nanowires in the high symmetry transport orientations. (a) [100], (b) [110] and (c) [111] transport orientation. The [110] nanowire in (b) is the one analyzed in this work. The width direction is [1-10] and the height direction is [001]. Equivalently, the top/bottom quantization surfaces are (001) (perpendicular to the [001] direction). The left/right quantization surfaces are (1-10) (perpendicular to the [1-10] direction).

Figure 2: Quantizations of different surfaces

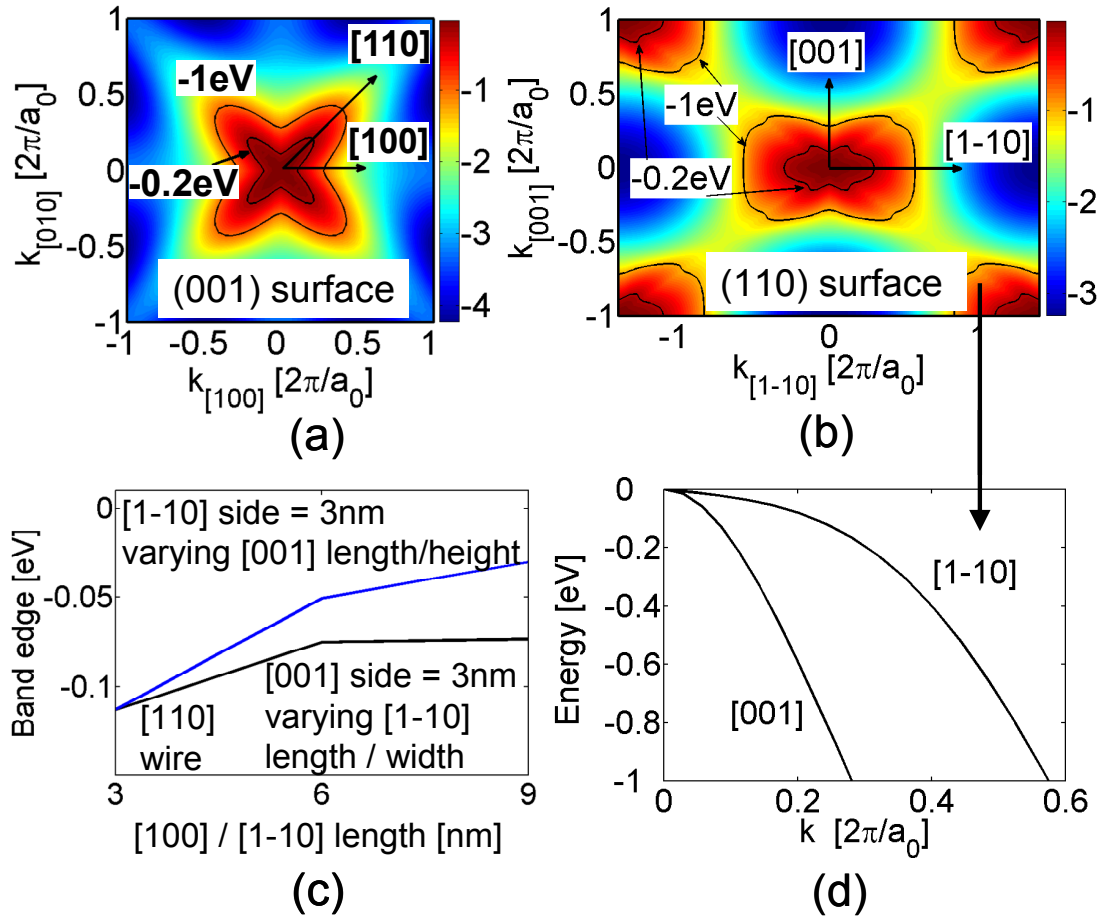

(a)

(b)

(c)

(d)

Figure 2 caption:

(a-b) k-space energy surface contours of the heavy hole calculated using the full 3D k-space information of the Si Brillouin zone. The energy contours for $E=-0.2$eV and $E=-1$eV are plotted. (a) The (001) surface. The anisotropy is evident between the [100] and [110] directions. (b) The (110) surface. (Or equivalently, -45° "cut" through the center of (a)). (c) The band edge of a [110] transport oriented nanowire for the cases: (1) The size of the [001] directed side increases from 3nm to 9nm while keeping the size of the [1-10] side at 3nm (blue). (2) The size of the [1-10] directed side increases from 3nm to 9nm while keeping the size of the [001] at 3nm (black). (d) The dispersions of the heavy-hole in the [1-10] and [001] directions.

Figure 3: Current surface under variation of the dimensions

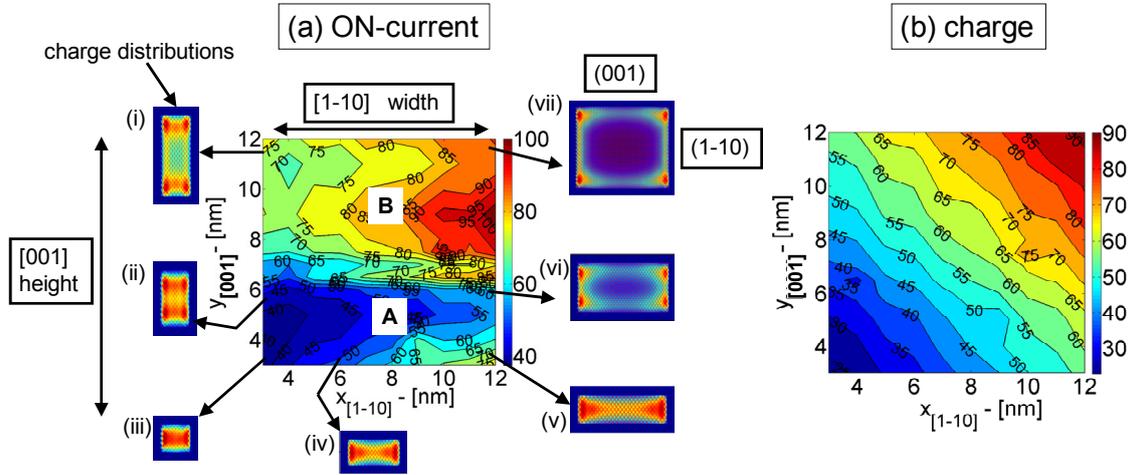

Figure 3 caption:

(a) The ON-current contour plot in μA as a function of the [110] nanowires' side size variations. The *x*-axis is the width in the [1-10] direction, and the *y*-axis is the height in the [001] direction. The side figures show the nanowires' cross sections and the charge distribution at high bias for the devices indicated by the arrows in the centered figure. Wires shown (width-[1-10] x height-[001]): (i) 3nm x 12nm. (ii) 3nm x 6nm. (iii) 3nm x 3nm. (iv) 6nm x 3nm. (v) 12nm x 3nm. (vi) 12nm x 6nm. (vii) 12nm x 12nm. The top/bottom surfaces are (001). The left/right surfaces are (1-10). (b) The total charge contour plot in the devices of (a). All parameters in the simulations are the same for all devices with only the sizes changing. The gate bias is $V_G$=1V and the drain bias $V_D$=0.5V.

Figure 4: Velocity surface and bands

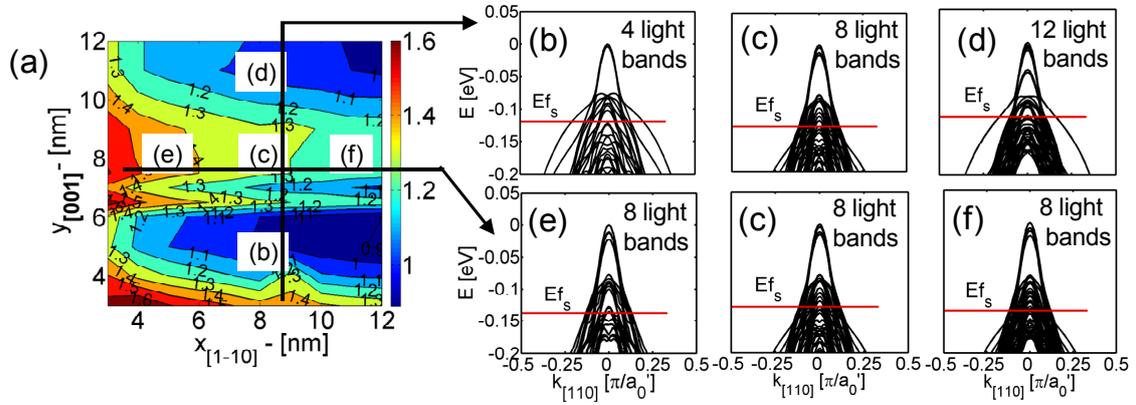

Figure 4 caption:

(a) The average velocity contour plot in the devices of Fig. 3(a). (b) The dispersion relation for the 8nm x 5nm wire (point b). (c) The dispersion relation for the 8nm x 8nm wire (point c). (d) The dispersion relation for the 8nm x 12nm wire (point d). (e) The dispersion relation for the 5nm x 8nm wire (point e). (f) The dispersion relation for the 12nm x 8nm wire (point f). $a_0'$ is the wires' unit cell length.

Figure 5: Strain engineering

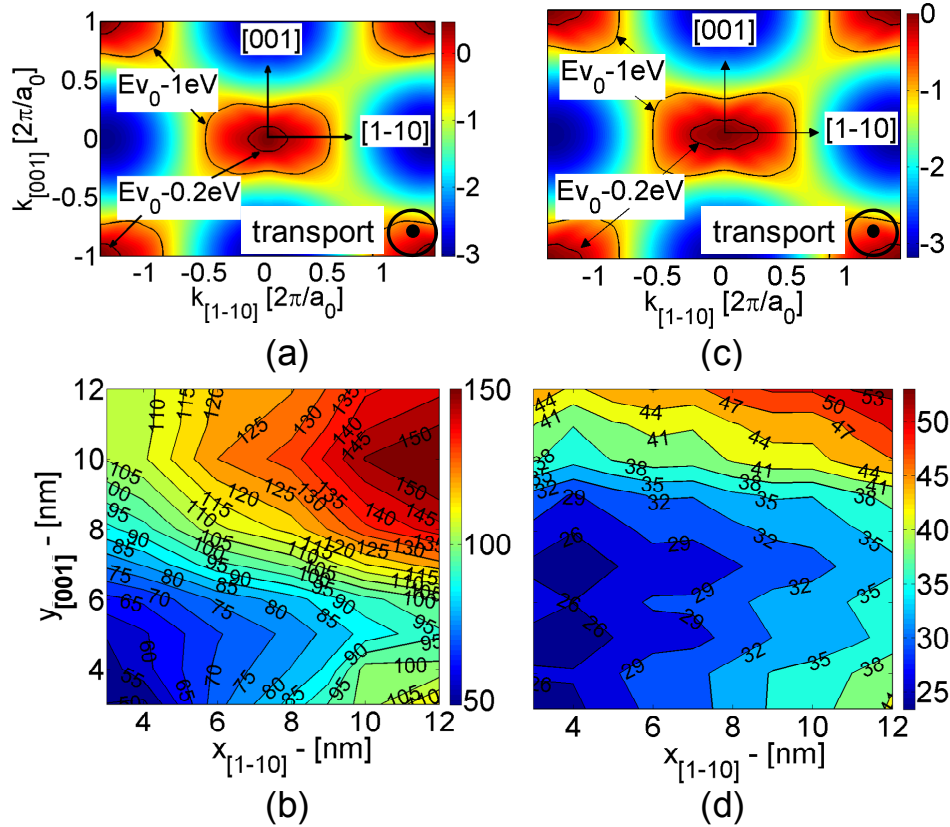

Figure 5 caption:

(a) (110) quantization energy surface contour of the heavy hole using 3% compressive strain in the transport [110] direction, 3% tensile strain in the [1-10] quantization direction and 0.05% compressive strain in the [100] quantization direction. The energy contours for $E$=-0.2eV and $E$=-1eV below the valence band maximum are plotted. Compared to Fig. 3a the quantization mass is more isotropic. (b) Same as Fig. 3a for the case of the strain tensor described above in (a). (c) (110) quantization energy surface contour of the heavy hole using 1% compressive strain in the transport [110] direction, 3% compressive strain in the [1-10] quantization direction and 3% compressive strain in the [001] quantization direction. The energy contours for $E$=-0.2eV and $E$=-1eV below the valence band maximum are plotted. Compared to Fig. 3a the quantization mass is more anisotropic. (d) Same as Fig. 3a for the case of the strain tensor described above in (c).